\documentclass[aps,preprint]{revtex4}%
\usepackage{amsfonts}
\usepackage{amsmath}
\usepackage{amssymb}
\usepackage{graphicx}
\usepackage{pdfpages}
\usepackage{float}
\usepackage{color}
\usepackage{xcolor}
\usepackage{caption}
\usepackage{lineno,hyperref}
\usepackage{changes}%
\hypersetup{colorlinks =true, allcolors = blue}
\providecommand{\U}[1]{\protect\rule{.1in}{.1in}}

\begin{document}
\begin{center}
{\Large Massless Dirac Perturbations of black holes in f(Q) gravity: quasinormal modes and weak deflection angle 
}

Ahmad Al-Badawi and Sohan Kumar Jha

Department of Physics, Al-Hussein Bin Talal University, P. O. Box: 20,
71111, Ma'an, Jordan.

Department of Physics, Chandernagore College, Chandernagore, Hooghly, West
Bengal, India

E-mail: ahmadbadawi@ahu.edu.jo, sohan00slg@gmail.com
\\
{\Large Abstract}
\end{center}
This article considers a static and spherical black hole (BH) in f(Q) gravity. f(Q) gravity is the extension of symmetric teleparallel general relativity, where both curvature and torsion are vanishing, and gravity is described by nonmetricity. In this study, we investigate the possible implications of quasinormal modes (QNM) modified Hawking spectra, and deflection angles generated by the model. The WKB method is used to solve the equations of motion for massless Dirac perturbation fields and explore the impact of the nonmetricity parameter ($Q_{0}$). Based on the QNMs computation, we can ensure that the BH is stable against massless Dirac perturbations and as $Q_{0}$ increases the the oscillatory frequency of the mode decrease. We then discuss the weak deflection angle in the
weak field limit approximation. We compute the deflection angle up to the fourth order of approximation and show how the nonmetricity parameter affects it. We find that the $Q_{0}$ parameter reduces the deflection angle.

\section{Introduction}
The gravitational effects can be manifested through three different avenues. In general relativity, the space-time curvature describes the gravity. In teleparallel and symmetric teleparallel theories of general relativity, the gravitational effects are ascribed to torsion and non-metricity, respectively. The symmetric teleparallel $f(Q)$ theory of gravity has been at the center of extensive study for quite some time now. With the help of data from various astrophysical observations such as Type Ia Supernovae (SNe Ia) and quasars, authors in their manuscript \cite{Q2} have analyzed the validity of the $f(Q)$ theory. The validity of $f(Q)$ cosmological models has also been investigated in the article \cite{Q3} through the embedding approach. The authors in \cite{Q4} have studied black holes (BHs) in the $f(Q)$ gravity. The static and spherically symmetric solutions resulting from the $f(Q)$ gravity immersed in an anisotropic have been elucidated in \cite{Q6}. The authors in their article \cite{Q7} have analyzed the wormhole geometries in the $f(Q)$ gravity where they employed the linear equation of state and anisotropic solutions. In \cite{Q8}, authors have elucidated the possibility of traversal wormholes consistent with energy conditions. Various studies regarding wormholes in the $f(Q)$ gravity have been done in \cite{Q9, Q11}. Authors in \cite{Q10} have scrutinized various aspects of static and spherically symmetric solutions in the $f(Q)$ gravity. \\     
Observations regarding various astrophysical phenomena provide an excellent avenue to probe different theories of gravity. One such phenomenon is the quasinormal modes (QNMs) of BHs. Since QNMs bear the imprints of the underlying space-time, they can be utilized to get important aspects of the space-time. These modes are referred to as quasinormal as, unlike normal modes, they are transient in nature. These modes represent oscillations of the BH that eventually die out owing to the emission of gravitational waves. QNMs are basically complex-valued numbers where the real part provides the frequency of emitted gravitational waves and the imaginary part gives the decay rate or the damping rate. BHs undergo three different phases after perturbation: inspiral, merger, and ringdown. QNMs are related to the ringdown phase for remnant BHs. A significant number of articles have delved into studying QNMs of various black holes [\citenum{CM}-\citenum{YY1}]. The author in articles \cite{skj, skj2} has studied QNMs of non-rotating loop quantum gravity and Simpson-Visser BHs, respectively. In article \cite{alif}, the authors have studied QNMs of a static and spherically symmetric BH in the $f(Q)$ gravity for scalar and electromagnetic. In this article, we intend to study QNMs of the BH for a massless Dirac field and investigate its time evolution, and we will not discuss the relevant material in References \cite{alif}.\\
Another astrophysical phenomenon that encodes important information of the underlying space-time is gravitational lensing. In the absence of any massive object, light rays travel in a straight line. But, in the presence of BHs, due to their strong gravitational fields, light rays get deflected. Thus, BHs act as gravitational lenses. Since the deflection angle is a function of different parameters that arise in the theory of gravity under consideration, studying gravitational lensing provides us with a way to analyze the effect of various theories of gravity on the observable. The phenomenon of gravitational lensing has been studied extensively in strong as well as weak field limits in various articles [\citenum{BOZZ}-\citenum{COSC2}] . Motivated by previous studies and with the intention to study the effect of the nonmetricity parameter on the deflection angle, we study the gravitational lensing in the $f(Q)$ gravity in the weak field limit.\\  
Our primary focus of this article is to gauge the impact of the nonmetricity scalar $Q_0$ on astrophysical observations such as QNMs and gravitational lensing. It is imperative to probe the signature of additional parameter(s) so that with future experimental results, we can either validate or invalidate the new solution. The structure of the paper is as follows: Sect. II briefly discusses the spacetime of a static and spherical BH in $f(Q)$ gravity. In Sect. III, we describe the massless Dirac perturbation at the
neighborhood of static BH in $f(Q)$ gravity.  In Sec. IV, the QNMs
frequencies are evaluated using the sixth-order WKB method. The time evolution profile of the Dirac perturbation is provided in Sec. V. In Sec. VI, we study the deflection angle in the weak field limit. We end our manuscript with a brief discussion of the results.

\section{A brief discussion on static BH in f(Q) gravity}
The action for the $f(Q)$ gravity is \cite{action1}
\begin{equation}
S=\int \sqrt{-g}\mathrm{d} x^{4}\left( \frac{1}{2}f\left( Q\right) +L_{m}\right) ,
\label{ac1}
\end{equation}
where $g$ is the determinant of the metric $g_{\mu \nu}$, $f(Q)$ is an arbitrary function of the nonmetricity $Q$ and $%
L_{m}$ is the matter Lagrangian density. Varying the action (\ref{ac1}) 
with respect to the metric $g_{\mu \nu}$ gives the field equation%
\begin{equation}
-T_{\mu \nu }=\frac{2}{\sqrt{-g}}\nabla _{\alpha }\left( \sqrt{-g}\frac{%
\partial f}{\partial Q}P_{\mu \nu }^{\alpha }\right) +\frac{1}{2}g_{\mu \nu}f+\frac{\partial f}{\partial Q}\left( P_{\mu \alpha \beta }Q_{\nu
}^{\alpha \beta }-2Q_{\alpha \beta \mu }P_{\nu }^{\alpha \beta }\right) .
\end{equation}%
The stress-energy momentum tensor for cosmic matter content is determined by%
\begin{equation}
T_{\mu \nu }\equiv \frac{-2}{\sqrt{-g}}\frac{\delta \left( \sqrt{-g}%
L_{m}\right) }{\delta g^{\mu \nu }}.
\end{equation}%
Varying Eq. (\ref{ac1}) with respect to the connection, one obtains%
\begin{equation}
\nabla \mu \nabla _{\nu }\left( \sqrt{-g}\frac{\partial f}{\partial Q}%
P_{\alpha }^{\mu \nu }\right) =0.
\end{equation}%
The metric ansatz for a generic static and spherically symmetric spacetime
is written as%
\begin{equation}
\mathrm{d} s^{2}=-e^{a\left( r\right) }\mathrm{d} t^{2}+e^{b\left( r\right) }\mathrm{d} r^{2}+r^{2}\left(
 \mathrm{d}\theta ^{2}+\sin ^{2}\theta \mathrm{d}\phi ^{2}\right) .
\end{equation}%
For this ansatz, the nonmetricity scalar $Q$ can be written as \cite{Wang}
\begin{equation}
Q\left( r\right) =\frac{2e^{b\left( r\right) }}{r}\left( a^{\prime }+\frac{1%
}{r}\right) ,
\end{equation}%
where a prime $\left( ^{\prime }\right) $  denotes a derivative with respect
to the radial coordinate $r$. For constant nonmetricity scalar $\left(
Q=Q_{0}\right) ,$ the above equation can be written as%
\begin{equation}
a^{\prime }\left( r\right) =-\frac{Q_{0}re^{b\left( r\right) }}{2}-\frac{1}{r%
}.  \label{q13}
\end{equation}%
Following \cite{Wang}, the components of the field equation in the case of vacuum can be
written as 
\begin{equation}
\frac{\partial f\left( Q_{0}\right) }{\partial Q}\frac{e^{-b}}{r}\left(
a^{\prime }+b^{\prime }\right) =0,
\end{equation}%
\begin{equation}
\frac{f\left( Q_{0}\right) }{2}+\frac{\partial f\left( Q_{0}\right) }{%
\partial Q}\left( Q_{0}+\frac{1}{r^{2}}\right) =0,  \label{q12}
\end{equation}%
\begin{equation}
\frac{\partial f\left( Q_{0}\right) }{\partial Q}\left[ \frac{Q_{0}}{2}+%
\frac{1}{r^{2}}+e^{-b}\left( \frac{a^{\prime \prime }}{2}+\left( \frac{%
a^{\prime }}{4}+\frac{1}{2r}\right) \left( a^{\prime }-b^{\prime }\right)
\right) \right] =0.
\end{equation}%
Equation  (\ref{q12}) implies that 
\begin{equation}
f\left( Q_{0}\right) =0,\qquad \frac{\partial f\left( Q_{0}\right) }{%
\partial Q}=0.
\end{equation}%
Those two conditions imply that 
\begin{equation}
f(Q)=\sum\limits_{n}a_{n}\left( Q-Q_{0}\right) ^{n}, \label{cn1}
\end{equation}%
where $a_{n}$ represents arbitrary model parameters. Assume $e^{a\left(
r\right) }=1-2M/r$  then Eq. (\ref{q13}) results%
\begin{equation}
e^{b}=\frac{-2}{Q_{0}r\left( r-2M\right) },
\end{equation}
The metric of the static spherical BH in $f(Q)$ gravity is given by \cite{Wang}
\begin{equation}
\mathrm{d}s^{2}=f(r)\mathrm{d}t^{2}-\frac{1}{g(r)}\mathrm{d}r^{2}-r^{2}\left( \mathrm{d}\theta ^{2}+\sin
^{2}\theta \mathrm{d}\phi ^{2}\right)  \label{M1}
\end{equation}%
where 
\begin{equation}
f\left( r\right) =1-\frac{2M}{r},\qquad g(r)=\frac{-Q_{0}r^{2}}{2}\left( 1-%
\frac{2M}{r}\right) ,  \label{mf1}
\end{equation}%
in which, $M$ is the BH mass, $Q_{0}$ is the constant nonmetricity scalar
which must have $Q_{0}<0$. In contrast to a Schwarzschild BH, this BH has a
nonmetricity scalar $Q_{0}$. Moreover, it modifies the scalars of metric (%
\ref{M1}) as follows: 
\begin{eqnarray}
R &=&\frac{3\left( r-M\right) Q_{0}r+2}{r^{2}}\rightarrow \text{Ricci scalar}
\label{scalar1} \\
R_{\mu \nu }R^{\mu \nu } &=&\frac{\left( 9M^{2}-14Mr+6r^{2}\right)
Q_{0}^{2}r^{2}+8\left( r-M\right) Q_{0}r+4}{2r^{4}}\rightarrow \text{Ricci
squared scalar,}  \notag \\
\mathcal{K} &=&\frac{\left( 9M^{2}-8Mr+3r^{2}\right) Q_{0}^{2}r^{2}+4\left(
r-2M\right) Q_{0}r+4}{r^{4}}\rightarrow \text{Kretchmann scalar.}  \notag
\end{eqnarray}%
According to Eq. (\ref{scalar1}), it is obvious that this BH has a physical
singularity at $r=0$, and that the nonmetricity scalar $Q_{0}$\ considerably
alters the scalars associated with this BH.

\begin{figure}
{{\includegraphics[width=7.5cm]{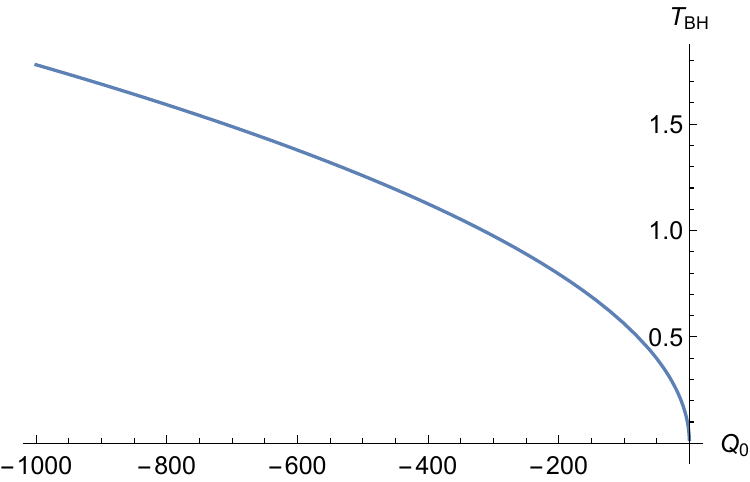} }}
    \caption{  Hawking temperature versus $Q_{0}$.} \label{fig1}
\end{figure}

The Hawking temperature of the metric (\ref{M1}) can be calculated using the
surface gravity given by \cite{wald}
\begin{equation}
\kappa =\nabla _{\mu }\chi ^{\mu }\nabla _{\nu }\chi ^{\nu },
\end{equation}%
where $\chi ^{\mu }$ is the timelike Killing vector field. Therefore, the
Hawking temperature of the static BH (\ref{M1}) reads%
\begin{equation}
T_{BH}=\frac{1}{4\pi \sqrt{-g_{tt}g_{rr}}}\left. \frac{dg_{tt}}{dr}%
\right\vert _{r=r_{h}}=\frac{1}{4\pi \sqrt{-\frac{2}{Q_{0}r^{2}}}}\frac{2M}{%
r^{2}}=\frac{1}{4\pi }\sqrt{-\frac{Q_{0}}{2}}.  \label{temp1}
\end{equation}%
From Eq. (\ref{temp1}), it is obvious that the nonmetricity scalar $Q_{0}$\
increases the Hawking temperature of the static BH. The behavior of the Hawking temperature is plotted on Fig. \ref{fig1}. 
 
\section{\ Massless Dirac perturbation }

In this section, we will discuss the massless Dirac perturbation in the  static and spherical BH in $f(Q)$ gravity. 
To study the massless spin- 1/2 field, we will use the Newman-Penrose formalism
\cite{NP}. The Dirac equations \cite{chand} are
given by
\begin{equation*}
\left( D+\epsilon -\rho \right) F_{1}+\left( \overline{\delta }+\pi -\alpha
\right) F_{2}=0,
\end{equation*}%
\begin{equation}
\left( \Delta +\mu -\gamma \right) F_{2}+\left( \delta +\beta -\tau \right)
F_{1}=0,  \label{Dirac1}
\end{equation}%
where $F_{1},$ $F_{2}$ represent the Dirac spinors and $D=l^{\mu }\partial
_{\mu },\Delta =n^{\mu }\partial _{\mu }$, $\delta =m^{\mu }\partial _{\mu }$
and $\overline{\delta }=\overline{m}^{\mu }\partial _{\mu }$\ are the
directional derivatives. The suitable choice for the null tetrad basis
vectors in terms of elements of the metric (\ref{M1}) is 
\begin{equation*}
l^{\mu }=\left( \frac{1}{f},\sqrt{\frac{g}{f}},0,0\right) ,\qquad n^{\mu }=
\frac{1}{2}\left( 1,-\sqrt{fg},0,0\right) ,
\end{equation*}%
\ 
\begin{equation}
m^{\mu }=\frac{1}{\sqrt{2}r}(0,0,1,\frac{i}{\sin \theta }),\qquad \overline{m%
}^{\mu }=\frac{1}{\sqrt{2}r}(0,0,1,\frac{-i}{\sin \theta }).  \label{7}
\end{equation}%
Based on those definitions, we find that the only non-vanishing components of
spin coefficients are: 
\begin{align}
\rho & =-\frac{1}{r}\sqrt{\frac{g}{f}},\qquad \mu =-\frac{\sqrt{fg}}{2r}%
,\qquad  \notag \\
\gamma & =\frac{f^{\prime }}{4}\sqrt{\frac{g}{f}},\qquad \beta =-\alpha =%
\frac{\cot \theta }{2\sqrt{2}r}.  \label{8}
\end{align}%
Decoupling the differential Eqs. (\ref{Dirac1}) yields a single equation of
motion for $F_{1}$ only (which is actually the equation of motion for a
massless Dirac field):%
\begin{equation}
\left[ \left( D-2\rho \right) \left( \Delta +\mu -\gamma \right) -\left(
\delta +\beta \right) \left( \overline{\delta }+\beta \right) \right]
F_{1}=0,  \label{Dirac2}
\end{equation}%
With Eqs. (\ref{7}) and (\ref{8}), we can express Eq. (\ref{Dirac2})
explicitly as

\begin{equation*}
\left[ \frac{1}{2f}\partial _{t}^{2}-\left( \frac{\sqrt{fg}}{2r}+\frac{%
f^{\prime }}{4}\sqrt{\frac{g}{f}}\right) \frac{1}{f}\partial _{t}-\frac{%
\sqrt{fg}}{2}\sqrt{\frac{g}{f}}\partial _{r}^{2}-\sqrt{\frac{g}{f}}\partial
_{r}\left( \frac{\sqrt{fg}}{2r}+\frac{f^{\prime }}{4}\sqrt{\frac{g}{f}}%
\right) \right] F_{1}
\end{equation*}%
\begin{equation}
+\left[ \frac{1}{\sin ^{2}\theta }\partial _{\phi }^{2}+\frac{i\cot \theta }{%
\sin \theta }\partial _{\phi }+\frac{1}{\sin \theta }\partial _{\theta
}\left( \sin \theta \partial _{\theta }\right) -\frac{1}{4}\cot ^{2}\theta +%
\frac{1}{2}\right] F_{1}=0  \label{Dirac3}
\end{equation}%
To decople Eqs. (\ref{Dirac3}) into  radial and angular parts, we consider the
spin- $1/2$ wave function in the form of 
\begin{equation}
F_{1}=R\left( r\right) A_{l,m}\left( \theta ,\phi \right) e^{-ikt}
\label{anz1}
\end{equation}%
where $k$ is the frequency of the incoming Dirac field and $m$ is the
azimuthal quantum number of the wave. Therefore, the angular part of Eq. 
 (\ref{Dirac3}) becomes
\begin{equation}
\left[ \frac{1}{\sin ^{2}\theta }\partial _{\phi }^{2}+\frac{i\cot \theta }{%
\sin \theta }\partial _{\phi }+\frac{1}{\sin \theta }\partial _{\theta
}\left( \sin \theta \partial _{\theta }\right) -\frac{1}{4}\cot ^{2}\theta +%
\frac{1}{2}+\lambda \right] A_{l,m}\left( \theta ,\phi \right) =0,
\end{equation}%
where $\lambda $ is the separation constant. While the radial part reads%
\begin{equation}
\left[ \frac{-k^{2}}{2f}-\left( \frac{\sqrt{fg}}{2r}+\frac{f^{\prime }}{4}%
\sqrt{\frac{g}{f}}\right) \frac{-ik}{f}-\frac{\sqrt{fg}}{2}\sqrt{\frac{g}{%
f}}\partial _{r}^{2}-\sqrt{\frac{g}{f}}\partial _{r}\left( \frac{\sqrt{fg}}{%
2r}+\frac{f^{\prime }}{4}\sqrt{\frac{g}{f}}\right) \right] R\left( r\right)
=0  \label{rad1}
\end{equation}%
Equation (\ref{rad1}) represents the radial Teukolsky equation for a
massless spin 1/2 field, which can be transformed into certain Schr\"{o}%
dinger-like wave equations:%
\begin{equation}
\frac{\mathrm{d}^{2}U_{\pm }}{\mathrm{d}r_{\ast }^{2}}+\left( k^{2}-V_{\pm }\right) U_{\pm }=0,
\label{master}
\end{equation}%
where the generalized tortoise coordinate $r_{\ast }$ is defined as $\frac{\mathrm{d}
}{\mathrm{d}r_{\ast }}=\sqrt{fg}\frac{\mathrm{d}}{\mathrm{d}r}$ and the potentials $V_{\pm }$ for the
massless spin 1/2 field are given by%
\begin{equation}
V_{\pm }=\frac{\left( l+1/2\right) ^{2}}{r^{2}}f\pm \frac{\left(
l+1/2\right) }{r}\sqrt{fg}\left( \frac{f^{\prime }}{2\sqrt{f}}-\frac{\sqrt{f}%
}{r}\right) .  \label{effp1}
\end{equation}
\begin{figure}
    \centering
{{\includegraphics[width=7.5cm]{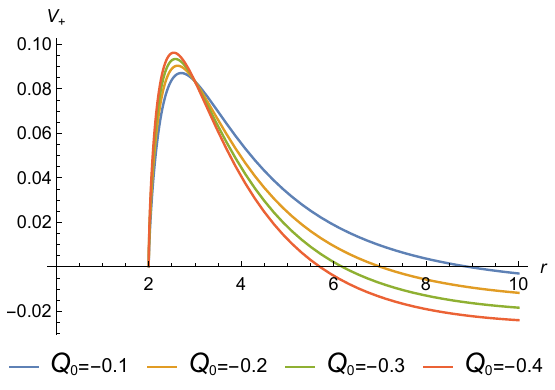} }}\qquad
    {{\includegraphics[width=7.5cm]{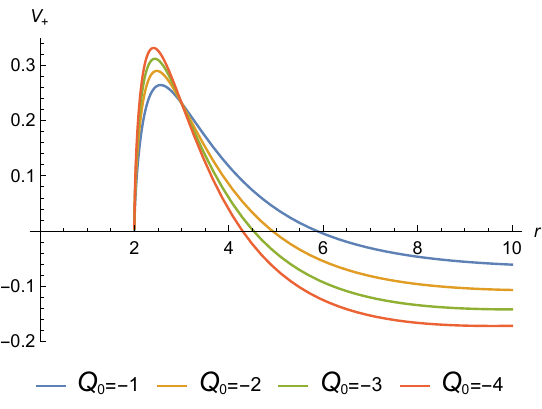}}}
    \caption{The variation of the potential (\ref{effp2}) w.r.t. $r$
for massless Dirac field with different values of $Q_{0}$.}
    \label{fig2}
\end{figure}

\section{QNM of a static BH in f(Q) gravity}

Using the master wave Eq. (\ref{master}) with the boundary condition of
purely outgoing waves at infinity and purely ingoing waves at the event
horizon, we can calculate the spectrum of the massless Dirac field's complex
frequencies, i.e. QNMs, for a static spherical BH (\ref{M1}). To estimate the QNMs of the BH considered in this study, we will use a well-established method known as the WKB method.  According to Ref. \cite{schutz}, Schutz and Will introduced the first-order WKB method or technique for the first time. Although this method can approximate QNM, its error is relatively higher. This is why higher-order WKB methods have been implemented in the study of BH QNMs, and we will apply the sixth order method in this study \cite{iyer,konoplya}. The formula for the complex QNM is reported in \cite{konoplya}. 

\begin{figure}
    \centering 
    \includegraphics [width=18cm]{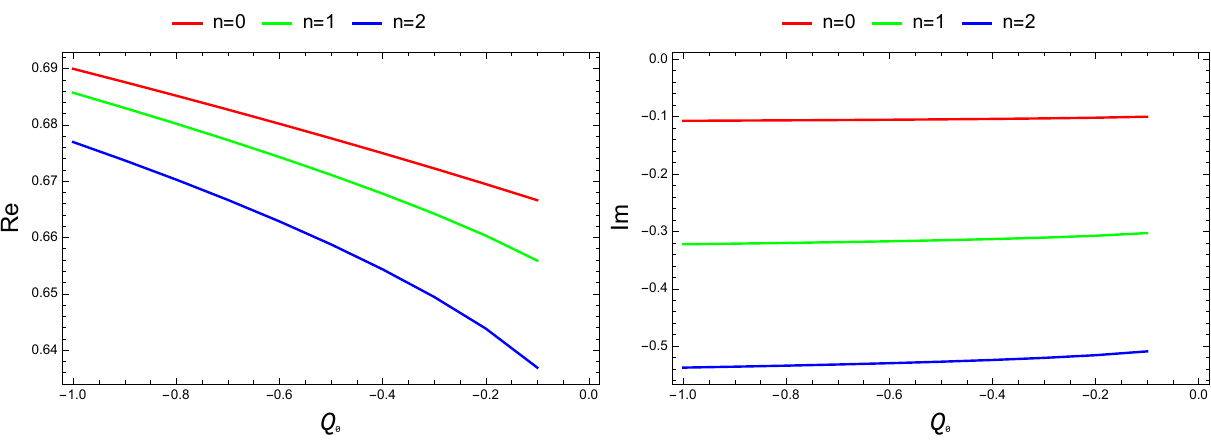}
    \caption{The QNM of the BH with $M=1$. }
    \label{fig3}
\end{figure}

\begin{figure}
    \centering
{{\includegraphics[width=7.5cm]{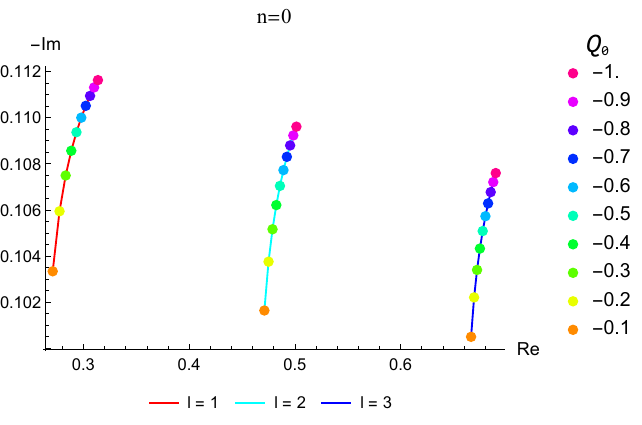} }}\qquad
    {{\includegraphics[width=7.5cm]{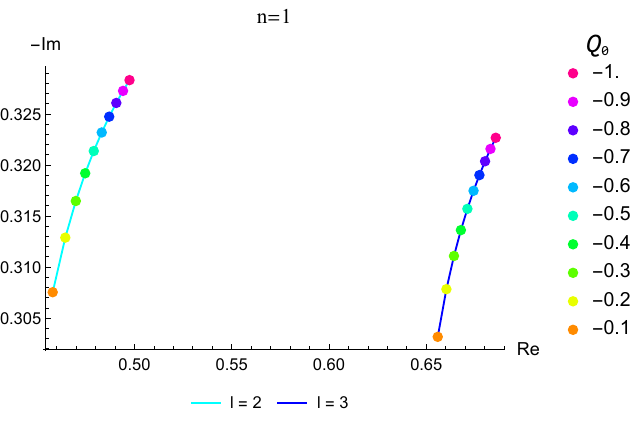}}}\qquad
    {{\includegraphics[width=7.5cm]{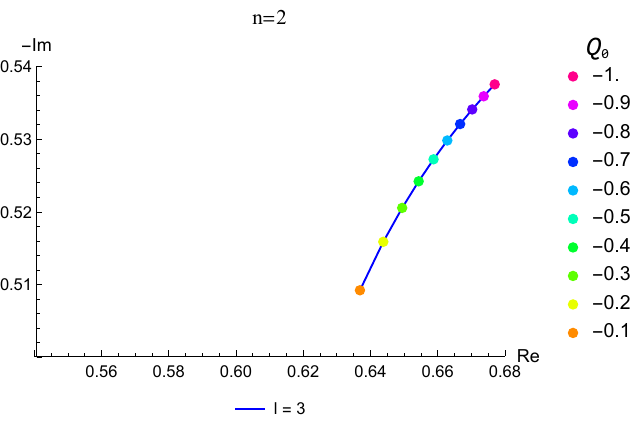}}}
    \caption{Complex frequency plane for $l=1, l=2$ and $l=3$ which shows the behavior of the quasinormal frequencies, Table \ref{tab1}.}
    \label{fig4}
\end{figure}

We can calculate the quasinormal frequencies of the massless Dirac field for
the static BH using the potential derived in the
previous section and given by Eq. (\ref{effp1}). We will select $V_{+}$ as
this potential.  Considering $V_{+}$ rather than $V_{-}$ is sufficient for an analogue analysis because $V_{-}$ behaves qualitatively similar to $V_{+}$.
Therefore, 
\begin{equation}
V_{+}=\frac{\left( l+1/2\right) ^{2}}{r^{2}}\left( 1-\frac{2M}{r}\right)
+\left( \frac{3M-r}{r^{3}}\right) \frac{\sqrt{\frac{-Q_{0}}{2}\left(
r-2M\right) ^{2}}}{\sqrt{1-\frac{2M}{r}}}\left( l+1/2\right) .  \label{effp2}
\end{equation}%
Let's examine briefly the behavior of the potential for the static BH
described above. By observing the behavior of the potential, one can gain
some insight into the QNMs. To understand how $Q_{0}$ impacts the effective potential, we plotted the potential in Fig. \ref{fig2} for smaller and larger values of $Q_{0}$.  It is seen from Fig.  \ref{fig2} that as $Q_{0}$ increases potential  decreases. This analysis demonstrates that the model parameter $Q_{0}$ has a significant influence on potential behaviour. This implies that the parameter $Q_{0}$ may have an effect on the QNM spectrum. 
In Table \ref{tab1}, we have shown the QNMs for massless Dirac  perturbation for different $l$ values with $M =
1$. Based on the results, it can be concluded that all frequencies have a negative imaginary part, confirming the stability of the BH modes found. We obtain Figures \ref{fig3} and \ref{fig4} to examine the effect of the nonmetricity scalar parameter $Q_{0}$ on QNM frequencies. Figure \ref{fig3} shows that as $Q_{0}$ increases the real part or the  oscillatory frequency of the mode decreases.  Figure \ref{fig4} indicates that an increase in $Q_{0}$ leads to a decrease in the real component and a decrease in the
imaginary component. Additionally, it is evident from the analysis of the imaginary part of the QNM frequencies that, with increasing $Q_{0}$, the damping rate modestly increases.  
\begin{table}[]
    \centering
\begin{tabular}{|c|c|c|c|c|}
\hline 
$Q_{0}$ & $n$ & $l=1$ & $l=2$ & $l=3$ \\ \hline \hline
$-0.1$ & 
\begin{tabular}{l}
$0$ \\ 
$1$ \\ 
$2$%
\end{tabular}
&
\begin{tabular}{l}
$0.270934-0.103337i$ \\ 
 \\  \\
\end{tabular}
& 
\begin{tabular}{l}
$0.471115-0.101636i$ \\ 
$0.457984-0.307536i$ \\ \\
\end{tabular}
& 
\begin{tabular}{l}
$0.666599-0.100494i$ \\ 
$0.655877-0.303166i$ \\ 
$0.636862-0.509225i$%
\end{tabular}
\\ \hline
$-0.2$ & 
\begin{tabular}{l}
$0$ \\ 
$1$ \\ 
$2$%
\end{tabular}
& 
\begin{tabular}{l}
$0.277386-0.105941i$ \\ 
\\ \\
\end{tabular}
& 
\begin{tabular}{l}
$0.475056-0.103754i$ \\ 
$0.464455-0.312882i$ \\ \\
\end{tabular}
& 
\begin{tabular}{l}
$0.669464-0.102204i$ \\ 
$0.660328-0.307832i$ \\ 
$0.643805-0.515889i$%
\end{tabular}
\\ \hline
$-0.3$ & 
\begin{tabular}{l}
$0$ \\ 
$1$ \\ 
$2$%
\end{tabular}
& 
\begin{tabular}{l}
$0.283195-0.107492i$ \\ 
\\ \\
\end{tabular}
& 
\begin{tabular}{l}
$0.478814-0.10516i$ \\ 
$0.469871-0.31647i$ \\ \\
\end{tabular}
& 
\begin{tabular}{l}
$0.672249-0.103396i$ \\ 
$0.664217-0.311084i$ \\ 
$0.649423-0.520552i$%
\end{tabular}
\\ \hline
$-0.4$ & 
\begin{tabular}{l}
$0$ \\ 
$1$ \\ 
$2$%
\end{tabular}
& 
\begin{tabular}{l}
$0.288498-0.108565i$ \\ 
\\ \\
\end{tabular}
& 
\begin{tabular}{l}
$0.482409-0.106209i$ \\ 
$0.474687-0.319185i$ \\ \\
\end{tabular}
& 
\begin{tabular}{l}
$0.67496-0.104323i$ \\ 
$0.667781-0.313618i$ \\ 
$0.654326-0.524206i$%
\end{tabular}
\\ \hline
$-0.5$ & 
\begin{tabular}{l}
$0$ \\ 
$1$ \\ 
$2$%
\end{tabular}
& 
\begin{tabular}{l}
$0.293395-0.109369i$ \\ 
\\ \\
\end{tabular}
& 
\begin{tabular}{l}
$0.485859-0.107041i$ \\ 
$0.47909-0.321365i$ \\ \\
\end{tabular}
& 
\begin{tabular}{l}
$0.677602-0.105082i$ \\ 
$0.671116-0.3157i$ \\ 
$0.658756-0.527231i$%
\end{tabular}
\\ \hline
$-0.6$ & 
\begin{tabular}{l}
$0$ \\ 
$1$ \\ 
$2$%
\end{tabular}
& 
\begin{tabular}{l}
$0.297954-0.110001i$ \\ 
\\ \\
\end{tabular}
& 
\begin{tabular}{l}
$0.489179-0.107725i$ \\ 
$0.483182-0.32318i$ \\ \\
\end{tabular}
& 
\begin{tabular}{l}
$0.68018-0.105725i$ \\ 
$0.674276-0.317469i$ \\ 
$0.662841-0.529818i$%
\end{tabular}
\\ \hline
$-0.7$ & 
\begin{tabular}{l}
$0$ \\ 
$1$ \\ 
$2$%
\end{tabular}
& 
\begin{tabular}{l}
$0.30223-0.110515i$ \\ 
\\ \\
\end{tabular}
& 
\begin{tabular}{l}
$0.492384-0.108302i$ \\ 
$0.487027-0.324728i$ \\ \\
\end{tabular}
& 
\begin{tabular}{l}
$0.682701-0.106282i$ \\ 
$0.677294-0.319005i$ \\ 
$0.666657-0.53208i$%
\end{tabular}
\\ \hline
$-0.8$ & 
\begin{tabular}{l}
$0$ \\ 
$1$ \\ 
$2$%
\end{tabular}
& 
\begin{tabular}{l}
$0.306261-0.110945i$ \\ 
\\ \\
\end{tabular}
& 
\begin{tabular}{l}
$0.495482-0.108797i$ \\ 
$0.490668-0.326072i$ \\ \\
\end{tabular}
& 
\begin{tabular}{l}
$0.685166-0.106771i$ \\ 
$0.680193-0.32036i$ \\ 
$0.670256-0.53409i$%
\end{tabular}
\\ \hline
$-0.9$ & 
\begin{tabular}{l}
$0$ \\ 
$1$ \\ 
$2$%
\end{tabular}
& 
\begin{tabular}{l}
$0.310081-0.111312i$ \\ 
\\ \\
\end{tabular}
& 
\begin{tabular}{l}
$0.498485-0.109229i$ \\ 
$0.494136-0.327255i$ \\ \\
\end{tabular}
& 
\begin{tabular}{l}
$0.687581-0.107206i$ \\ 
$0.682989-0.321571i$ \\ 
$0.673674-0.535897i$%
\end{tabular}
\\ \hline
$-1$ & 
\begin{tabular}{l}
$0$ \\ 
$1$ \\ 
$2$%
\end{tabular}
& 
\begin{tabular}{l}
$0.313714-0.111629i$ \\ 
\\ \\
\end{tabular}
& 
\begin{tabular}{l}
$0.501398-0.10961i$ \\ 
$0.497454-0.328308i$ \\ \\
\end{tabular}
& 
\begin{tabular}{l}
$0.689948-0.107597i$ \\ 
$0.685695-0.322663i$ \\ 
$0.676937-0.537537i$
\end{tabular}
\\ \hline 
\end{tabular} 
\caption{The QNM of the BH with $M=1$. }
    \label{tab1}
    \end{table}
    
\section{Ringdown waveform}
\begin{figure}
    \centering
{{\includegraphics[width=7.5cm]{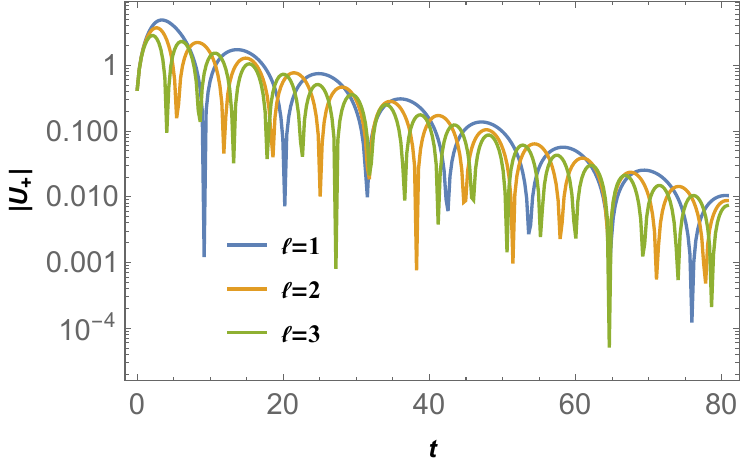} }}\qquad
    {{\includegraphics[width=7.5cm]{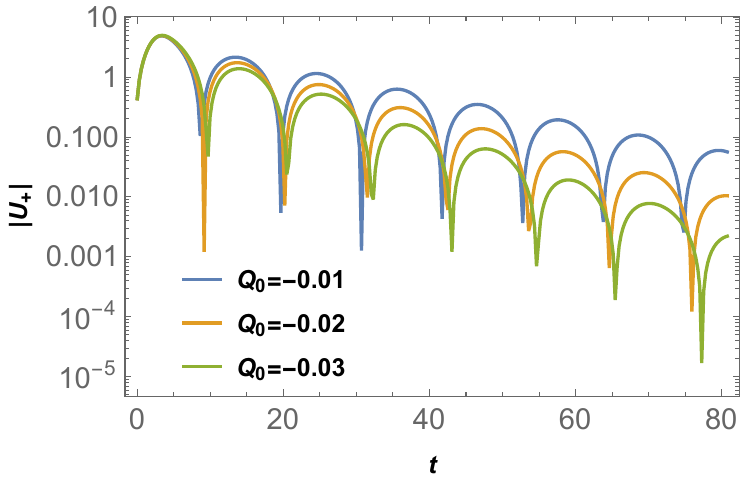}}}\caption{Time domain profile for massless Dirac field. The left one is for various values of nonmetricity parameter $Q_0$ with $\ell=1$ and the right one is for different multipole numbers with $Q_0=-0.02$.}
\label{ringing}
\end{figure}

In this section, we intend to study the effect of the nonmetricity scalar on the ringdown waveform of a massless Dirac field. To this end, we employ the time domain integration method elaborated in  \cite{gundlach1}. The initial conditions used here are:
\begin{equation}
    U_{+}(r_*,t) = \exp \left[ -\dfrac{(r_*-\hat{r}_{*})^2}{2\sigma^2} \right] \quad \text{and} \quad U_{+}(r_*,t)\vert_{t<0} = 0,  
\end{equation}
where $\hat{r}$ and $2\sigma^2$ are taken to be $0.4$ and $25$, respectively. The values of $\Delta t$ and $\Delta r_{*}$ are taken in order to satisfy the Von Neumann stability condition, $\frac{\Delta t}{\Delta r_*} < 1$.\\
In the left panel of Fig. \ref{ringing}, we provide the ringdown waveform for various values of the parameter $Q_0$ keeping $\ell=1$ and in the right panel, the waveform for various values of $\ell$, keeping $Q_0=-0.02$, is shown.
From Fig. \ref{ringing} we observe that the frequency of quasinormal modes increases as we increase the multipole number or decrease the nonmetricity parameter. On the other hand, we can clearly conclude from the figure that the decay rate decreases with $\ell$ but increases with a decrease in $Q_0$. These conclusions are in agreement with those drawn from the Table \ref{tab1}. Our study conclusively shows the significant impact the nonmetricity parameter has on quasinormal modes and time profile evolution of massless Dirac field.

\section{Deflection angle of black hole in nonplasma medium}

For calculating the deflection angle $\alpha $, we first assume a static,
spherically symmetric spacetime 
\begin{equation}
\mathrm{d}s^{2}=A(r)\mathrm{d}t^{2}-B\left( r\right) \mathrm{d}r^{2}-C\left( r\right) \left( \mathrm{d}\theta
^{2}+\sin ^{2}\theta \mathrm{d}\phi ^{2}\right)  \label{mf2}
\end{equation}%
The deflection angle $\alpha $ is given by \cite{arthur, weinb}
\begin{equation}
\alpha \left( r_{0}\right) =2\int_{r_{0}}^{\infty }\frac{1}{C}\sqrt{\frac{AB%
}{\frac{1}{b^{2}}-\frac{A}{C}}}dr-\pi ,  \label{ang1}
\end{equation}%
where $r_{0}$ is the light ray's distance of closest approach to the lens
and $b$ is the impact parameter given by 
\begin{equation}
b=\sqrt{\frac{C\left( r\right) }{A\left( r\right) }}.
\end{equation}%
For some simple cases, the above integral can only be solved analytically.
As a result, Keeton and Petters \cite{keeton} proposed that this result can be
approximated by a series of the form%
\begin{equation}
\alpha \left( b\right) =a_{0}+a_{1}\left( \frac{M}{b}\right) +a_{2}\left( 
\frac{M}{b}\right) ^{2}+a_{3}\left( \frac{M}{b}\right) ^{3}+O\left( \frac{M}{%
b}\right) ^{4},  \label{ang2}
\end{equation}
where $a_{i}$ are coefficients to be found. To apply this method, let us
substitute the functions 
\begin{equation}
A(r)=1-\frac{2M}{r},B(r)=\frac{-2}{Q_{0}r^{2}}\left( 1-\frac{2M}{r}\right)
^{-1},C\left( r\right) =r^{2},
\end{equation}%
in Eq. (\ref{ang1}), therefore the deflection angle becomes
\begin{equation}
  \alpha \left( r_{0}\right) =2\int_{r_{0}}^{\infty }\frac{\sqrt{2}}{r^{2}}%
\sqrt{\frac{-b^{2}r}{Q_{0}\left( r^{3}+b^{2}\left( 2M-r\right) \right) }}%
\mathrm{d} r-\pi   \label{ang3}
\end{equation}
Let us assume the following coordinates transformation, namely 
\begin{equation}
x=\frac{r_{0}}{r},h=\frac{M}{r_{0}},
\end{equation}%
Then, the impact parameter becomes%
\begin{equation}
b=\frac{r_{0}}{\sqrt{1-2\frac{M}{r_{0}}}}.  \label{imp1}
\end{equation}
Therefore Eq. (\ref{ang3}) becomes

\begin{equation}
\alpha \left( r_{0}\right) =2\sqrt{2}\int_{0}^{1}\sqrt{-\frac{x^{2}}{%
Q_{0}\left( x^{2}-1-2h\left( x^{3}-1\right) \right) }}dx-\pi ,
\end{equation}%
Assuming the weak field regime $\left( h<<1\right) ,$ then the integrand is
now expanded into a Taylor series, and the integral is solved term by term,
giving us%
\begin{equation}
\alpha \left( r_{0}\right) =-\frac{2\sqrt{2}}{\sqrt{-Q_{0}}}+\frac{2\sqrt{2}%
}{\sqrt{-Q_{0}}}h-\frac{3\sqrt{2}\left( \pi -5\right) }{\sqrt{-Q_{0}}}h^{2}+%
\frac{7\sqrt{2}\left( 45\pi -112\right) }{16\sqrt{-Q_{0}}}h^{3}+O\left(
h\right) ^{4}.  \label{alpa12}
\end{equation}%
This expression Eq. (\ref{alpa12}) is, however, coordinate-dependent for the
reason that it refers to the distance of closest approach. By using Eq. (\ref%
{imp1}), then we can write%
\begin{equation}
r_{0}=4\frac{M}{b}+\frac{15\pi }{4}\left( \frac{M}{b}\right) ^{2}+\frac{128}{%
3}\left( \frac{M}{b}\right) ^{3}+\frac{3465\pi }{64}\left( \frac{M}{b}%
\right) ^{4}+O\left( h\right) ^{5}.
\end{equation}%
At the end, we can write the deflection angle Eq. (\ref{alpa12}) in terms of 
$M/b$ as follows%
\begin{equation}
\alpha \left( b\right) =\frac{\sqrt{2}}{\sqrt{-Q_{0}}}+\frac{-4+3\pi }{2%
\sqrt{2}\sqrt{-Q_{0}}}\left( \frac{M}{b}\right) +\frac{26-3\pi }{2\sqrt{2}%
\sqrt{-Q_{0}}}\left( \frac{M}{b}\right) ^{2}+\frac{-384+279\pi }{16\sqrt{2}%
\sqrt{-Q_{0}}}\left( \frac{M}{b}\right) ^{3}+O\left( \frac{M}{b}\right) ^{4}.
\label{alpha99}
\end{equation}%
It is obvious that the nonmetricity scalar constant $Q_{0}$ decreases the
deflection angle. In order to discuss the effect of impact parameter ($b$)
on deflection angle, we plot Eq. (\ref{alpha99}) for different values of $%
Q_{0}$. Figure shows that the deflection angle decreases with impact
parameter for all values of $Q_{0}$ and remains positive.
\begin{figure}
    \centering
    \includegraphics{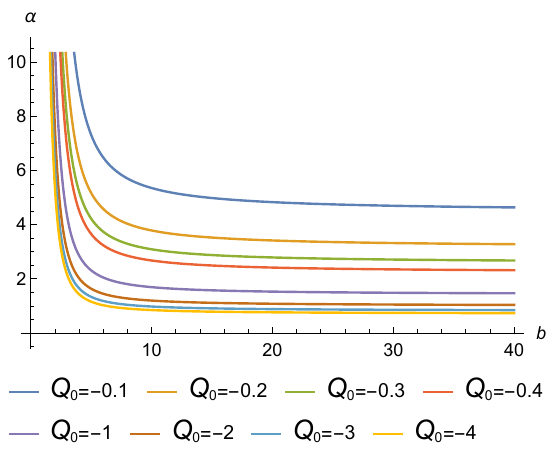}
    \caption{The behavior of deflection angle $\alpha$ Eq. (\ref{alpha99}) with respect to impact parameter for varying
$Q_{0}$}
    \label{fig5}
\end{figure}

\section{Conclusion}

In this work, we have considered  a static and spherical BH in $f(Q)$ gravity. The $f(Q)$ gravity is an extension of symmetric teleparallel general relativity, where both curvature and torsion vanish, and gravity is explained by nonmetric terms. We have studied the QNMs of a
of the massless Dirac
field. We used the 6th order WKB method to perform our calculations. We have determined how the nonmetricity  parameter influence the potential and the real and imaginary parts of quasinormal frequencies. It is found that, as $Q_{0}$ increases potential decreases. Moreover,  the nonmetricity parameter has a greater influence on QNMs for real  values of QNMs than for imaginary values, as shown in Fig. \ref{fig3}. According to Fig. \ref{fig4}, an increase in $Q_{0}$ leads to a decrease in the real component and a decrease in the imaginary component. As a result of the analysis of the real part and imaginary part of the QNM frequencies, it is evident that the oscillatory frequency of the mode decreases as $Q_{0}$ increases, while the damping rate increases modestly with $Q_{0}$.
The results obtained from the time domain analysis are in agreement with those obtained from the numerical analysis. It is necessary to understand the theory of $f(Q)$ by studying quasinormal modes from BHs as one of the interesting and widely studied properties of a perturbed BH spacetime.  As a result of our study, we demonstrate that the BH solution studied can generate significantly different quasinormal modes than a Schwarzschild BH \cite{sch1}. Furthermore, the authors of \cite{newq1} investigated the quasinormal modes of a test massless scalar field around static black hole solutions in $f(T)$ gravity. The authors investigate how the $f(T)$ model parameter $\alpha$ and orbital angular momentum l affect field decay behaviour.  It is discovered that when $l$ increases, the period of quasinormal vibration decreases, mimicking the Schwarzschild scenario. Furthermore, the field decay behaviour varies smoothly for varied $\alpha$. Interestingly, our findings are identical to those in $f(T)$ gravity. 

In addition, we calculated the deflection angle using the Keeton and Petters method, which is an approximation of gravitational lensing. We discovered post-post-Newtonian metric coefficients by comparing the expanded metric function to the standard post-post-Newtonian metric. Later, we determined the bending angle coefficients and compared them with the general form of the Schwarzschild metric to obtain the final results shown in Eq.(\ref{alpha99}). Moreover, the authors of \cite{newq2, newq3} have derived a tight constraint upon the parameter space of $f(T)$ theory by virtue of galaxy-galaxy weak lensing surveys. In this work, they use galaxy-galaxy weak gravitational lensing to obtain more precise restrictions on hypothetical departures from General Relativity. Using $f(T)$ gravitational theories to quantify the deviation, we discovered that the quadratic correction on top of General Relativity is preferred. As shown in \cite{newq4}, this method can be generalized to the strong gravitational lensing case around compact astronomical objects. In \cite{newq4} they explore gravitational lensing effects within the framework of $f(T)$ gravity, focusing on the  Singular Isothermal Sphere and the Singular Isothermal Ellipsoid mass models. Their findings show that under $f(T)$ gravity, for both mass models, the deflection angle is higher than in General Relativity.  In our article, we have considered static and spherically symmetric BH, whereas a BH can have charge and rotation. The presence of charge and rotation will impact astrophysical observations quantitatively. However, the qualitative impact of the nonmetricity scalar $Q_0$ on astrophysical observations will remain the same for fixed values of charge and rotation as static and spherically symmetric BHs can be considered as a limiting case of a rotating, charged BH where the rotation parameter and charge are zero. \\ Finally, As an application of $f(Q)$ gravity, we have examined the theoretical implications for $f(Q)$ cosmology in terms of model construction. In general, the situation $Q = Q_0$ can play the function of the cosmological constant in Schwarzschild-like solutions, implying a cosmological model of $f(Q)$ gravity for Dark Energy. Eq. (\ref{cn1}) illustrates one possible functional form for $f(Q)$. By setting the coefficients, one can create customised models of $f(Q)$ gravity. The most basic example is the polynomial of Q, which was developed and studied as a cosmological model of $f(Q)$ gravity \cite{newq87}.

\end{document}